\documentclass[fleqn,10pt]{wlscirep}
\usepackage[utf8]{inputenc}
\usepackage[T1]{fontenc}
\usepackage[page,title]{appendix}
\usepackage{url}
\usepackage{graphicx}
\usepackage{amssymb}
\usepackage{bm}
\usepackage{mathrsfs}
\usepackage{amsmath}
\usepackage{amsfonts}
\usepackage{color}
\usepackage{xspace}
\usepackage{tikz}
\usepackage{slashed}
\usepackage{cancel}
\usepackage{hyperref}
\usepackage[utf8]{inputenc}  
\usepackage{xcolor}
\usepackage{slashed}
\usepackage{fancyhdr}
\usepackage{rotating}
\usepackage{multirow}
\usepackage{xcolor,colortbl}
\usepackage[sort, numbers]{natbib}
\usepackage{graphicx}
\usepackage[freestanding]{hepunits}
\usepackage{cleveref}
\usepackage{tocloft}
\usepackage{enumitem}
\usepackage{tcolorbox}

\begin{document}


\title{ 
ACE Science Workshop Report
}

\author[1]{Stefania Gori (ed.)}
\author[2]{Nhan~Tran (ed.)}
\author[3]{Karri~DiPetrillo}
\author[4]{Bertrand Echenard}
\author[2]{Jeffrey Eldred}
\author[2]{Roni Harnik}
\author[2]{Pedro Machado}
\author[2]{Matthew Toups}

\author[2]{Robert Bernstein}
\author[2,6]{Innes Bigaran}
\author[7]{Cari Cesarotti}
\author[8]{Bhaskar Dutta}
\author[2]{Christian Herwig}
\author[9]{Yonatan Kahn}
\author[2]{Sergo Jindariani}
\author[4]{Ryan Plestid}
\author[10]{Vladimir Shiltsev}
\author[5]{Matthew Solt}
\author[11]{Alexandre Sousa}
\author[2]{Diktys Stratakis}
\author[6]{Zahra Tabrizi}
\author[5]{Anil Thapa}
\author[2]{Jacob Zettlemoyer}
\author[11]{Jure Zupan}

\affil[1]{University of California Santa Cruz, Santa Cruz, CA 95064, USA}
\affil[2]{Fermi National Accelerator Laboratory, Batavia, IL 60510, USA}
\affil[3]{University of Chicago, Chicago, IL 60637, USA}
\affil[4]{California Institute of Technology, Pasadena, CA 91125, USA}
\affil[5]{University of Virginia, Charlottesville, VA 22904, USA}
\affil[6]{Northwestern University, Evanston, IL 60208, USA}
\affil[7]{Massachusetts Institute of Technology, Cambridge, MA 02139, USA}
\affil[8]{Mitchell Institute and Texas A\&M University, College Station, TX 77843, USA}
\affil[9]{University of Illinois Urbana Champaign, Urbana, IL 61801, USA}
\affil[10]{Northern Illinois University, DeKalb, IL 60115, USA}
\affil[11]{University of Cincinnati, Cincinnati, OH 45221, USA}


\begin{abstract}
We summarize the Fermilab Accelerator Complex Evolution (ACE) Science Workshop, held on June 14-15, 2023. The workshop presented the  strategy for the ACE program in two phases: ACE Main Injector Ramp and Target (MIRT) upgrade and ACE Booster Replacement (BR) upgrade.  Four plenary sessions covered the primary experimental physics thrusts: Muon Collider, Neutrinos, Charged Lepton Flavor Violation, and Dark Sectors.  Additional physics and technology ideas were presented from the community that could expand or augment the ACE science program.  Given the physics framing, a parallel session at the workshop was dedicated to discussing priorities for accelerator R\&D.  Finally, physics discussion sessions concluded the workshop where experts from the different experimental physics thrusts were brought together to begin understanding the synergies between the different physics drivers and technologies.  
 In December of 2023, the P5 report was released setting the physics priorities for the field in the next decade and beyond, and identified ACE as an important component of the future US accelerator-based program. Given the presentations and discussions at the ACE Science Workshop and the findings of the P5 report, we lay out the topics for study to determine the physics priorities and design goals of the Fermilab ACE project in the near-term.  
\end{abstract}

\maketitle


\clearpage

\clearpage
\tableofcontents

\clearpage
\section{Executive Summary}
The Fermilab Accelerator Complex Evolution (ACE) Science Workshop~\cite{ACEworkshop}, held on June 14-15, 2023, brought together an international cohort of physicists with expertise in theory, experiment, and accelerator technologies to begin a coordinated discussion defining the Fermilab ACE physics program. The aim of this intensive workshop was to review the motivations and current state of the primary physics thrusts and their accelerator needs and to outline a strategic road map to understand the complementarity of their emerging scientific demands.
A detailed description of the preliminary ACE approach is outlined in the Proton Intensity Upgrade - Central Design Group (PIU-CDG) report~\cite{PIU-CDGReport}.

\vspace{0.2cm}
\noindent ACE consists of two primary phases: 
\begin{enumerate}
    \item \textbf{ACE-MIRT (Main Injector Ramp and Target) upgrade}: this aims to upgrade the Main Injector to reduce the ramp time and deliver more beam power to DUNE (max $\sim$ 2.1\,MW) as soon as possible. This also requires target R\&D to ensure that DUNE can handle up to 2.4\,MW of beam power, the ultimate design goal.
    \item \textbf{ACE-BR (Booster Replacement) upgrade}:  the Booster replacement aims to replace the Booster synchrotron which accelerates protons from 800\,MeV to 8\,GeV to \textbf{deliver} the full $>$2.4\,MW of beam power to DUNE, and to \textbf{enable} the development of the next generation of US accelerator-based particle physics experiments while modernizing the complex to provide \textbf{reliable} beam to all its users.
\end{enumerate}

\vspace{0.2cm}
\noindent The scientific potential of the Fermilab ACE science program is very broad and includes a wide variety of physics topics discussed during the Snowmass 2021 process~\cite{Butler:2023eah} across the energy, neutrino, rare processes and precision, theory, and cosmic frontiers. 
The primary experimental physics thrusts and related experiments presented at the workshop were:
\begin{itemize}[noitemsep]
    \item \textbf{Charged Lepton Flavor Violation (CLFV) experiments}: using high intensity beams of muons aim to significantly improve indirect searches for new physics up to scales of $\mathcal O(10^5$\,TeV). 
    \item \textbf{Dark sector experiments}: using high intensity beams of protons and muons aim to explore uncovered parameter space of thermal dark matter models, as well as models that address the strong CP problem (axion-like-particles), the origin of neutrino masses (sterile neutrinos), and the hierarchy problem (dark scalars).
    \item \textbf{Muon Collider} (MC): enables direct searches for new physics at the $\mathcal{O}$(10\,TeV) scale with intense proton beams that can serve as technology demonstrators and a front-end proton driver facility. It also enables unique precision measurements of Standard Model (SM) particles, like the Higgs boson.
    \item \textbf{Neutrino experiments beyond DUNE}: enable exploration of the neutrino sector beyond DUNE via direct searches connected to anomalies and high precision Standard Model measurements.
\end{itemize}
While these focus areas comprise the primary experimental physics thrusts considered, other experimental proposals or ideas outside of these thrusts were presented in open sessions for remarks from the community, as well as in parallel discussion sessions. For example, rare meson decays, mono-energetic neutrino beams, or muon beams could be used to test dark sector physics. ACE can also be used for spin physics studies and energy research.  The breadth of these ideas can be synergistic with the primary experimental physics thrusts or expand the overall physics potential of ACE.

The synergy between the several physics thrusts was also discussed during the workshop. Neutrino detectors and neutrino beam measurements can be used to test the physics of the dark sector. A muon collider could be used to test non-standard neutrino interactions and dark matter freeze-out  models across a broad range of energies: high mass ($>$1\,TeV) WIMPs could be produced at the main interaction point, and light muon-philic dark sectors could be produced 
in the muon dumps. 
Neutrinos are a probe of lepton flavor symmetries, and, therefore, CLFV and neutrino experiments provide complementary
probes of new physics. Finally, CLFV experiments can explore the parameter space of new sub-GeV dark particles with flavor violating couplings.

\subsection*{Outcomes}
    
The preliminary points for study and outcomes of the workshop are discussed below. We discuss the ACE project design constraints, technology R\&D, and physics synergies:

\subsection*{ACE design parameters}
\vspace{0.2cm}
\begin{itemize}[noitemsep]
    \item \textbf{ACE-MIRT (Main Injector Ramp and Targetry) era: }
        \begin{itemize}[noitemsep]
        \item The 8\,GeV proton intensity and accelerator timeline economics during the ACE-MIRT era needs to be understood in more detail because the Main Injector (MI) cycle time reduction provides less beam power for the muon program, short baseline neutrino program, and MC R\&D.
        \item The ACE program should continue to provide a fraction of MI cycles to the 120\,GeV beam slow extraction program to support test beam, dark sector physics, and spin physics.
        \item An accumulator ring at 0.8\,GeV for pulsed proton beam would greatly enhance the physics capabilities of the proton beam dump dark sector program and the CLFV program. 
        \end{itemize}
    \item \textbf{ACE-BR (Booster Replacement) -- 2\,GeV Linac:} A 2\,GeV Linac is a part of all the current ACE-BR designs. The benefits and trade-offs of an earlier construction start need to be understood:
        \begin{itemize}[noitemsep]
        \item Due to increased muon production vs. proton energy, a 2\,GeV beam would improve the physics reach of next generation CLFV experiments.
        \item A 2\,GeV CW beam could serve as good MC cooling demonstrator front-end. Its benefits versus the 8\,GeV booster beam during the ACE-MIRT era require more study.
        \item Design of an accumulator ring optimized for 2\,GeV but that can operate at 0.8\,GeV should be explored to enable flexible operations across the ACE-MIRT and ACE-BR eras.
        \item Resource benefits by beginning the 2\,GeV Linac construction soon after PIP-II completion needs to be understood. 
        \end{itemize}
    \item \textbf{ACE-BR (Booster Replacement) -- 2$\rightarrow$8\,GeV: }
        \begin{itemize}[noitemsep]
        \item To deliver 2.4\,MW to DUNE, each of the 6 ACE-BR configurations are sufficient.
        \item To enable ACE to serve as a Muon Collider proton driver front-end, the ACE-BR configurations need to be re-optimized.
        \item An H$^-$ Linac option for ACE-BR is a more natural fit for MC than an RCS due to the requirement synergies with DUNE.
        \item Further study is needed to understand if an 8\,GeV accumulator ring used in a ACE-BR Linac scenario for DUNE would also be sufficient to serve as the MC proton driver accumulator ring and provide beam for other physics scenarios.
        \end{itemize}
\end{itemize}

\subsection*{Technology R\&D and Physics complementarity}
\vspace{0.2cm}
\begin{itemize}[noitemsep]
    \item \textbf{Accelerator R\&D: }
        \begin{itemize}[noitemsep]
        \item High Power Targetry is the highest priority accelerator R\&D area because it is required during the ACE-MIRT era to achieve the DUNE physics goals. It is also identified as a key area of synergistic R\&D for the MC and a future CLFV program. 
        \item Other high priority accelerator R\&D topic areas include high power H$^-$ stripping (including laser stripping), MC cooling demonstrator, and high current accumulator ring designs that can operate at 0.8-2 GeV.
        \end{itemize}
    \item \textbf{Physics synergies to be studied:}
        \begin{itemize}[noitemsep]
        \item The physics case for a neutrino factory needs to be re-visited in a post-DUNE scenario. A number of important points were discussed, especially tau neutrino physics and precision neutrino measurements, but the physics cases need to be collated.
        \item Better understanding of how neutrino mass benchmark models map onto CLFV physics reach would provide sharp targets for the CLFV program. 
        \end{itemize}
    \end{itemize}


\subsection*{Outlook and next steps}

Following the release of the P5 report~\cite{P5report}, the particle physics community has been provided with a roadmap outlining the physics priorities for the upcoming decade and beyond. The P5 report specifically highlights ACE as an important component of this plan (Recommendation 4(g), Recommendation 6, Area Recommendation 12), due to the potential presented by the Muon Collider. A primary aim of ACE is to facilitate the development of a proton driver essential for the Muon Collider while being able to deliver a world-class experimental physics program that can address several of the P5 physics drivers: Elucidate the Mysteries of Neutrinos; Reveal the Secrets of the Higgs Boson; Search for Direct Evidence of New Particles; Pursue Quantum Imprints of New Phenomena; Determine the Nature of Dark Matter. Defining the exciting potential of the ACE experimental program and the design of the facility are important recommendations of the P5 report.

The immediate future involves holding a series of workshops to gather input from the community and build consensus with the initial focus on refining the accelerator design to align with new P5 priorities. It is also necessary to understand how the various proposed physics concepts can fit together within the ACE design, including research and development paths. After the workshop focused on accelerator design, another physics workshop will be convened. With a concrete ACE layout and design, it will set out to establish physics priorities and explore the possibilities that will lead to a conceptual design (CD-0) in the following years to be presented to a targeted panel that determines the path forward for the US accelerator-based physics program~\cite{P5report}.

\clearpage
\section{Fermilab Accelerator Complex Evolution (ACE) and physics thrusts}
\subsection{Fermilab Accelerator Complex Evolution}

The Fermilab Accelerator Complex Evolution (ACE) consists of two primary phases, the Main Injector Ramp and Target (MIRT) upgrade and the Booster Replacement (BR) upgrade.  ACE-MIRT aims to upgrade the Main Injector to reduce the ramp time and deliver more beam power to DUNE (max $\sim$ 2.1\,MW) as soon as possible; this also requires target R\&D to ensure that DUNE can handle up to 2.4\,MW of beam power.  ACE-BR aims to replace the Booster synchrotron which accelerates protons from 800\,MeV to 8\,GeV to enable full delivery of $>$2.4\,MW of beam power to DUNE and \textbf{enable} the development of the next generation of US accelerator-based particle physics experiments, including multi-TeV collider research, while modernizing the complex to provide \textbf{reliable} beam to all its users.

{\textbf{The ACE-MIRT}} upgrade aims to significantly enhance the proton output directed towards the DUNE Phase I detector. By reducing the Main Injector (MI) cycle time from 1.2--1.4 seconds to 0.65 seconds, the initiative seeks to increase beam power and upgrade target systems to accommodate up to 2.1\,MW. Achieving this shortened cycle requires a substantial increase in voltage and electrical power, necessitating enhancements to power supplies, transformers, feeders, the size of service buildings, the introduction of additional cooling, and more tunnel penetrations. Additionally, the RF accelerating system will undergo modifications, either by replacing existing cavities with a new design that offers more volts per cavity or by adding more cavities of the current design, alongside updates to regulation, control, and instrumentation systems.

Addressing challenges of beam dynamics, losses, and shielding is crucial for ACE-MIRT. Upgrades to the MI collimators and the abort line are essential to accommodate the changes. During the shortened 0.65-second cycle, before the implementation of a Booster Replacement, the Recycler Ring will not be available for Mu2e, and only one Booster batch will be available for the 8\,GeV proton beam experiments. The target research and development (R\&D) segment focuses on identifying suitable candidate materials, conducting high-energy proton irradiation and pulsed-beam experiments to simulate expected radiation damage and beam interaction conditions, followed by Post-Irradiation Examination (PIE) to assess material properties, microscopic structural changes, and high-cycle fatigue testing.

Following ACE-MIRT, {\textbf{the ACE-BR upgrade}} aims to construct a new Booster to serve as a reliable platform for the future of the Accelerator Complex. This upgrade will continue to ensure the delivery of high-intensity beams for DUNE while enhancing the Fermilab's capability to support next generation accelerator experiments from precision measurements to searches for new physics with beams ranging from 1-120 GeV.  The project also intends to supply the high-intensity proton source necessary for future multi-TeV accelerator research, marking a significant leap forward in the complex's capabilities.

Exploring Booster replacement options, the project considers extending the Superconducting RF (SRF) Linac to higher energies or constructing a new Rapid-Cycling Synchrotron (RCS). Through the evaluation of three representative options for each approach, all requiring an extension of the SRF Linac to 2 GeV, the RCS option emerges as advantageous due to reduced space charge at the increased energy. Conversely, the high-energy linac option necessitates a beam with approximately 2\,GeV to leverage high-frequency, $\beta=1$, high-gradient cavities, which can be efficiently grouped and powered by a single, high-power klystron. The final selection and optimization of ACE-BR design will be based on outcomes of upcoming workshops and community design studies and physics prioritization.

Within this phased upgrade approach for the Fermilab accelerator complex, we can conceive \textbf{physics spigots} where additional experiments beyond DUNE would be feasible.  We divide them into the ACE-MIRT and ACE-BR eras.    

\begin{itemize}
    \item ACE-MIRT era:
    \begin{itemize}[noitemsep,topsep=0pt]
        \item \textbf{S0A}: 800\,MeV, PIP-II continuous wave
        \item \textbf{S0B}: 800\,MeV, PIP-II pulsed linac
        \item \textbf{S0C}: 800\,MeV, PIP-II pulsed with accumulator ring
        \item \textbf{S0D}: 8\,GeV, Booster pulsed
        \item \textbf{S0E}: 8\,GeV, Recycler Ring (RR) \& Delivery Ring (RR) (muon campus)
        \item \textbf{S0F}: 120\,GeV, MI continuous wave via slow extraction
    \end{itemize}
    \item ACE-BR era:
    \begin{itemize}[noitemsep,topsep=0pt]
        \item \textbf{S1A}: 2\,GeV, continuous wave or pulsed linac
        \item \textbf{S1B}: 2\,GeV, pulsed with accumulator ring
        \item \textbf{S1D}: 8\,GeV, pulsed, muon campus (RR/DR)
        \item \textbf{S1E}: 8\,GeV, Recycler Ring (RR) \& Delivery Ring (RR) (muon campus) [same as ACE-MIRT]
        \item \textbf{S1F}: 120\,GeV, MI continuous wave via slow extraction [same as ACE-MIRT]
    \end{itemize}
\end{itemize}



\subsection{Physics drivers and experimental thrusts}
ACE physics aligns well with all three science themes presented in the P5 report: (1) Decipher
the Quantum Realm; (2) Explore
New Paradigms in Physics; (3) Illuminate the Hidden Universe. Within each theme P5
identified two physics drivers, that represent the most promising avenues
of investigation for the next 10 to 20 years. ACE will address five out of these six physics drivers: {\bf{(1.1)}} Elucidate the Mysteries
of Neutrinos; {\bf{(1.2)}} Reveal the Secrets of
the Higgs Boson; {\bf{(2.1)}} Search for Direct Evidence
of New Particles; {\bf{(2.2)}} Pursue Quantum Imprints
of New Phenomena; {\bf{(3.1)}} Determine the Nature
of Dark Matter. In the following, we briefly highlight the main role of ACE in pursuing these goals.  

{\bf{(1.1)}} Parameters of the PMNS matrix are much less known than the corresponding parameters of the CKM matrix. Several models predict relations between the two. Therefore, an important goal will be to reach a similar level of precision. Future post-DUNE neutrino experiments with neutrinos produced from muon and anti-muon decays will allow the measurement of new oscillation observables through $\nu_e\to \nu_\mu$ and $\nu_e\to \nu_\tau$ transitions. 
{\bf{(1.2)}} A future high-energy muon collider will be able to achieve the most precise measurements of several Higgs properties including the $W W H H$ and $HHH$ couplings.
{\bf{(2.1)}} Several open questions in particle physics could be addressed by heavy NP particles with a mass beyond the LHC reach or by light NP particles that are only very weakly coupled to the SM (the ``dark sector''). A high-energy muon collider will be able to probe new particles with a mass above 10 TeV. Future neutrino experiments, low energy muon experiments, and proton fixed target experiments will be able to produce and detect a broad range of dark sector particles with a mass at or below the GeV scale. {\bf{(2.2)}}
Lepton flavor universality (LFU) and lepton flavor number are two approximate symmetries of the SM. Searches for charged lepton flavor universality breaking and for charged lepton flavor violation (CLFV) are therefore paramount. A well balanced program should include tests of $\mu \to e$, $\tau\to e$ and $\tau\to \mu$ transitions. Exciting prospects to test CLFV at Fermilab are expected. In particular, the Advanced Muon Facility will push the bound on the New Physics scale responsible for $\mu\to e$ transitions to $\sim 10^5$ TeV, a factor of several larger than what is obtainable at the Mu2e experiment, and help elucidate the nature of new physics in case of an observation. This is complementary to what can be done at (present and future) high-energy colliders searching for the Higgs decay $H\to \mu e$.
{\bf{(3.1)}} The progress in DM direct and indirect detection experiments, together with LHC searches, have put under tension vanilla models for Weakly-Interacting-Massive-particle (WIMP) DM. However, several WIMP scenarios, including the famous Higgsino DM scenario, remain unprobed. A high-energy muon collider will be able to probe these scenarios for heavy DM up to masses above 10 TeV. DM could belong to a light dark sector of particles only feebly interacting with the SM. To achieve the observed relic abundance for a DM particle with a mass below the few GeV scale, additional dark sector particles are generically necessary. 
Fermilab high-intensity beams offer a broad set of opportunities to test the dark sector paradigm, from DM searches in missing momentum signatures at muon fixed target experiments to visible dark sector particle (dark photons, axions, ...) searches at proton beam dump experiments to the production of dark sector particles from meson decays at neutrino experiments. 

Inspired by the above P5 physics drivers, we map them onto four primary experimental physics thrusts which are well-suited to the Fermilab ACE and around which we organize a future experimental program.  Four sessions at the ACE Science Workshop focused on these experimental thrusts and are summarized below.  

\subsubsection{Muon Collider}

The Muon Collider physics case is broad and exciting, including studying properties of the Higgs boson (including detailed characterization of the Higgs potential); rigorously testing predictions of the Standard Model (SM); investigating WIMP (Weakly Interacting Massive Particles) dark matter; exploring previously uncharted territories of quantum field theories; and enhancing research in the neutrino sector. Furthermore, the Muon Collider offers the potential for auxiliary fixed target and beam dump experiments to maximize its physics potential.  

The path to eventual realization of a Muon Collider lies through a vigorous R\&D program aimed at demonstrating key accelerator and detector technologies. In the workshop, R\&D challenges for both the Muon Collider detectors and accelerator development were presented and discussed. On the accelerator side, the R\&D plan includes development of a proton source compatible with the ACE plan, simulation of the proton accumulation and compression stages, and conducting experiments at existing facilities to demonstrate that the required proton bunch structure is attainable. The R\&D plan also involves studies of high-power target materials and designing a comprehensive target station that includes proton beam delivery, a production solenoid, cooling systems, a beam dump, and a particle selection chicane. Experimental R\&D studies of the target system would strongly benefit from the ACE beam. Beyond the proton driver and the target, ACE can offer a basis for a Muon Cooling Demonstrator facility that will allow the testing of key technologies related to ionization cooling. Such facility will consist of a muon source, a sequence of cooling cells, the upstream and downstream beam diagnostics instrumentation, and the associated infrastructure. The scheme will benchmark a realistic cooling lattice to 
give us the input, knowledge, and experience to design a real, buildable cooling channel for a Muon Collider. Such a demonstrator will require operation with proton beam, which ACE can potentially provide.

On the detector side, efforts to mitigate Beam-Induced Background (BIB) have demonstrated promising results through the application of innovative detector designs, advanced technologies (e.g. 5D detectors), and modern reconstruction software. The path forward involves substantial  R\&D on the promising detector technologies ( e.g. LGADs, MAPs, Dual Readout Calorimeters) and potential synergies within the high energy physics community in development of the next generation of advanced reconstruction algorithms. 

A key outcome of the workshop was that an 8\,GeV linac provides more flexible design over an RCS for the Muon Collider proton driver.  Another important outcome is that of the 6 configurations considered for ACE currently, all would need to be re-optimized to satisfy the needs of the Muon Collider proton driver.  

\subsubsection{Neutrino physics}

At neutrino factories, the precise knowledge of muon energy and charge allows for highly accurate neutrino energy spectra and exceptionally clean neutrino beams. Charge-identifying detectors can effectively eliminate beam-related background, enabling high-energy neutrino beams to facilitate measurements of electron neutrino to muon neutrino and electron neutrino to tau neutrino oscillations, thereby opening up new oscillation channels and the potential for additional observables. Precision neutrino oscillation measurements are crucial for probing new phenomena, and with current projects like DUNE and Hyper-K underway, muon storage rings emerge as a promising next step. As the Muon Collider garners renewed interest, the neutrino factory concept could serve as an interim step or a complementary program, contingent on a solid physics case and community support.

Fermilab is currently hosting a leading accelerator-based short-baseline neutrino program, focusing on neutrino-argon interaction measurements, neutrino flavor conversion, and dark sector physics. This effort is supported by the Booster Neutrino Beam facility, which has exceeded performance expectations and benefited from decades of data collection by various experiments (MiniBooNE, SciBooNE, ANNIE, MicroBooNE, and the SBN program), resulting in a well-characterized beam. Future upgrades to the Booster and extended running post-2027 could significantly enhance the scientific contributions of this program. The discovery of a new physics signal within the SBN program would be transformative, prompting further investigation either through additional neutrino or antineutrino running. Exploring the feasibility of such additional runs, especially in light of potential Booster upgrades and in conjunction with the forthcoming SBN results, along with possible detector enhancements, is highly valuable for the field.

\subsubsection{Charged Lepton Flavor Violation}
Charged Lepton Flavor Violating (CLFV) processes are critical in the search for new physics, offering both a high scale reach and the ability to diagnose models by identifying the source(s) of flavor breaking. The study of these processes is highly complementary to searches performed at colliders. If new physics is discovered at the LHC or other experiments, CLFV can provide exclusive insights into its symmetry structure. Conversely, if new physics remains elusive, CLFV stands as one of the best methods to investigate the mass scale of such undiscovered phenomena. Thanks to the availability of intense sources and their relatively long lifetime, muons offer a promising avenue to search for these reactions. A global experimental program is underway in the US, Europe, and Asia, with anticipated improvement in sensitivity of several orders of magnitude during this decade. In particular, the Mu2e experiment~\cite{bartoszek2015mu2e} at Fermilab is expected to probe muon-to-electron conversion with discovery potential at $\sim 10^{-16}$, a factor $10^4$ better than current limits. 

Beamline upgrades and a new facility at Fermilab could further extend the discovery potential by orders of magnitude, a tremendous opportunity for the muon program. The Mu2e-II project~\cite{Mu2e-II:2022blh} is a near-term evolution of Mu2e, planning to increase its reach by an order of magnitude. The Advanced Muon Facility (AMF)~\cite{CGroup:2022tli} is a more ambitious proposal for a new high-intensity muon science complex to enable broad muon science - including CLFV searches in at least four muon modes - with unprecedented sensitivity. The AMF complex is based on a fixed-field alternating gradient synchrotron (FFA) to create a cold, intense muon beam with low momentum dispersion. The FFA requires intense proton pulses with a bunch length of the order of 10 ns.  A compressor is needed to rebunch the PIP-II beam to obtain the desired structure; several options have been laid out in the context of the ACE upgrade. This facility would provide a unique opportunity to fully exploit the capabilities of PIP-II to produce a world-class physics program. Its development also has strong synergies with current R\&D efforts on a muon collider or a future dark matter program at FNAL, two key science priorities. 


\subsubsection{Accelerator-based dark sectors}

Accelerators are key tools for probing dark sectors in the laboratory, particularly within the thermal dark matter regime, and they represent a strategic direction for new physics research at intensity frontier experiments. There is a robust theoretical and experimental push to explore the most plausible models of dark sectors. 

The Fermilab Accelerator Complex is slated for upgrades to transition into the PIP-II era by the end of the decade, offering opportunities for additional enhancements as part of the ACE plan, such as incorporating an accumulator ring with short-pulse structures. These upgrades will open up significant physics opportunities, especially at a PIP-II beam dump facility that can accommodate various detector types with different thresholds in a dedicated experimental hall (see also Ref.~\cite{aguilararevalo2023physics}). Furthermore, existing and planned neutrino facilities are capable of searching for various dark sectors, often in tandem with their primary neutrino research objectives, across different energy ranges (800 MeV, 8 GeV, 120 GeV) that complement each other. Finally, the 120 GeV proton fixed target experiments provide an additional opportunity to searches for dark sectors that are sensitive to a broad spectrum of models. The proposed DarkQuest upgrade \cite{Apyan:2022tsd} to the existing SpinQuest experiment is envisioned as a catalyst for a sustained dark sector research program, leveraging existing infrastructure for new explorations in a short timeframe and with potential for further enhancements in detector and beam technology to increase sensitivity.

\subsection{Additional technology and physics directions}

The workshop included contributions from the community that spanned additional physics ideas, accelerator R\&D, and detector concepts that should be considered in the design of ACE. These could be provide synergy with the initially identified physics thrusts or may provide additional important  motivations or directions for the future facility.  

\subsubsection{Physics experiments}
\paragraph{KPIPE [8\,GeV]}
The KPIPE project aims to employ an extensive liquid scintillator detector to investigate the disappearance of $\nu_\mu$ neutrinos over a considerable distance, focusing on 236 MeV kaon decay-at-rest (KDAR) $\nu_\mu$ CC events. It relies on a pure and mono-energetic flux of muon neutrinos for its observations. The extended length of the detector is crucial for accurately measuring the oscillation wave. Notably, KPIPE demonstrates strong sensitivity, particularly effective in scenarios with high-$\delta m^2$ values. It serves as a valuable addition to the SBN program, offering complementary insights. Additionally, KPIPE is distinguished by its cost-effectiveness. The 8\,GeV proton beam requirements for KPIPE are high power and low duty factor, typically around $10^{-5}$.

\paragraph{FerMINI [120\,GeV]}

The FerMINI at MINOS experiment focuses on the exploration of millicharged particles (mCP) through a fixed-target setup employing stacks of scintillators. It investigates triple or double coincident signatures as indicators of the presence of mCPs.
There are several motivations behind the study of mCPs including the concept of charge quantization within the framework of Grand Unified Theories (GUTs) and string compactifications; a direct connection to models involving dark photons; and a connection to the 21 \,cm absorption spectrum, offering insights into fundamental aspects of cosmology and astrophysics.

\paragraph{LongQuest [120\,GeV]}
LongQuest, operating within the current NM4 SpinQuest facility, is dedicated to the pursuit of long-lived particle searches. This initiative has several advantages and distinctive features. Firstly, it offers superior shielding, ensuring minimal interference with ongoing SpinQuest operations. Additionally, the facility could also serve as an alternative site for the FerMINI project.
One of the key components of LongQuest is its dark photon decay fiducial volume. Positioned at a baseline distance ranging from 33 to 37\,m, this volume is specifically designed for the measurement of di-electrons or di-photons. Equipped with additional EMCal (Electromagnetic Calorimeter) and pre-Shower Detectors, this setup is sensitive to a range of dark sector physics scenarios with an expected $10^{18}$-$10^{20}$ Protons on Target (POT).

\paragraph{REDTOP [$\leq$\,2\,GeV]}

The REDTOP (Rare Eta Decays To Probe New Physics) experiment is being proposed, with the intent of collecting a data sample of order $10^{14}$ $\eta$ ($10^{12}$ $\eta'$) mesons for studying very rare decays. Such statistics are sufficient for investigating several symmetry violations, and for searching for particles and fields beyond the Standard Model. Utilizing a high intensity proton beam of $\leq$\,2\,GeV,  REDTOP will have sensitivity to processes that couple the Standard Model to New Physics through all four of the so-called \emph{portals}: the Vector, the Scalar, the Axion and the Heavy Lepton portal. The sensitivity of the experiment is also adequate for probing several conservation laws, in particular CP, T and Lepton Universality, and for the determination of the $\eta$ form factors, which is crucial for the interpretation of the recent measurement of muon $g-2$.

\paragraph{Muon Beam Dumps [$\leq$\,2\,GeV or 8\,GeV]}

Muon-beam dump experiments serve as excellent platforms for the exploration of dark scalars and other light dark sector particles, primarily due to their focus on muon couplings and their sensitivity to displaced decays. These experiments also offer the advantage of seamless integration with other muon-related research endeavors. 
Notably, the investigation of muonic dark sectors presents a promising avenue for addressing anomalies associated with muons, such as the muon g-2 anomaly. FNAL$\mu$, a prominent example, can be feasibly implemented within the existing infrastructure of Fermilab with minor modifications or additions, further enhancing its appeal and potential for significant scientific discovery.

\paragraph{DAMSA [$\leq$\,2\,GeV]}

DAMSA (Dump produced Aboriginal Matter Search at an Accelerator)~\cite{PhysRevD.107.L031901} is a very short baseline beam-dump experiment aimed at uncovering Dark Sector Particles (DSPs) using high-intensity, low-energy proton beams. Developed for 600 MeV proton beams initially, DAMSA is now adaptable to the 800 MeV PIP-II and ACE beams, offering compatibility with Fermilab's infrastructure. With collaborative efforts from ten US institutions and eight South Korean institutions, DAMSA  presents an exciting opportunity to transform Fermilab's accelerator facilities into world-leading DSP research hub.

\paragraph{Muonium and future muon physics [$\leq$\,2\,GeV]}

Both the muonium–antimuonium (M-$\bar{\rm{M}} $ (M is a bound $\mu e $ atom) and Mu2e experiments are critical and complementary in rare and precision physics searches. While muonium–antimuonium spectroscopy's previous results are over 20 years old, indicating a ripe opportunity for new efforts, the exploration of muonium gravity (testing antigravity on the $\mu^+$) has never been feasible until now, with the development of new techniques like SFHe production. 
The Muon Test Area (MTA) presents a unique opportunity to kickstart a leading Fermilab muonium program, facilitating initial investigations into muonium–antimuonium oscillations, spectroscopy, and gravity, each offering unique insights into new physics phenomena. These experiments hold promise for shedding light on CLFV physics and providing a novel avenue for testing gravitational couplings in the second generation.

\paragraph{Muon fixed target missing momentum [120\,GeV]}

The M$^3$ experiment utilizes the muon missing momentum technique to search for dark matter. Unlike LDMX, which employs electron beams, M$^3$ employs a muon fixed target experiment, necessitating adjustments to the detector, including a thicker target.
The experiment requires a low current muon beam with individual muons having energies greater than or equal to 10 GeV. Additionally, the muons must be individually identifiable, with an expected range of 10$^{10}$ to 10$^{13}$ Muons on Target (MOT).
Phase 1 (10$^{10}$ MOT) focuses on achieving complete coverage of the g-2 region, enabling the detection of any invisibly-decaying particle lighter than the muon. Phase 2 (10$^{13}$ MOT) allows for the exploration of large portions of the well-motivated dark matter parameter space with an increased MOT count.

\paragraph{Accelerator driven nuclear reactors [$\leq$\,2\,GeV]}

Mu*STAR, short for Muons' Subcritical Technology Advanced Reactor, is an accelerator-driven subcritical reactor (ADSR) designed to burn spent nuclear fuel from other reactors to produce carbon-free nuclear energy. It offers a viable solution to nuclear waste disposal while meeting public and legal demands for responsible nuclear technology.


\paragraph{Exploration of TMDs at a SpinQuest upgrade together with the Dark sector Physics [120\,GeV]}

The extension of the upcoming SpinQuest (E1039) experiment aims to extract gluon transversity distributions \cite{Keller:2022abm} from the Deuteron target (ND3). The prospective advancements of the Fermilab's Accelerator Complex Evolution (ACE) plan, enable not only the extraction of gluon transversity distributions with greater statistical accuracy with a deuteron target but also with a range of tensor-polarized nuclear targets with spin $\geq$ 1 to explore the nuclear dependences. The high-intensity beam and optimized timing between proton spills facilitate the uniqueness of Fermilab's capability for precise polarized target asymmetry measurements in Drell-Yan scattering. Also, this approach opens up the opportunity to measure ten additional leading twist quark transverse-momentum-dependent distributions (TMDs) for tensor-polarized targets, which have not been previously investigated well. The investigation will primarily focus on studying these TMDs through the Drell-Yan process, providing valuable insights into the nuclear EMC effect, as well as the 3-dimensional structure of the nucleon, supported by lattice QCD insights into nucleon spin components, marks a significant advance in understanding strong force, color confinement, and partonic interactions, setting a new standard for spin physics research. Additionally, the investigations into the Dark sector physics \cite{Apyan:2022tsd} will also be conducted at the NM4 facility, concurrently on the 120 GeV Main Injector, representing the proposed SpinQuest-upgrade.


\paragraph{Dedicated Muon EDMs [$\leq$\,2\,GeV]}

A potential EDM experiment could be conducted in the g-2 storage ring by modifying the quadrupole system to generate a radial electric dipole field pointing inward, with minimal changes to the current setup, except for altered orbit curvature and a higher potential difference across the inner/outer plates to create the electric field. The adaptation would allow for a significant reduction in the requirements for the existing magnetic dipole field, inflector, and kicker systems.
This setup presents an excellent chance to explore the systematics for a future dedicated run, marking the first application of the frozen spin technique and potentially enabling physics EDM measurements within a limited timeframe. Such an experiment could serve as a valuable proof of concept for a future dedicated EDM physics experiment.

\subsubsection{Technology development}
\paragraph{Test Beams: FTBF [120\,GeV] and ITA [800\,MeV]}

The Fermilab Test Beam Facility (FTBF) plays a critical role in supporting a broad spectrum of research and detector development with its two main beamlines, MTest and MCenter, offering particles ranging from 120 GeV protons to secondaries of approximately 200 MeV. Additionally, the facility includes an Irradiation Test Area (ITA) that provides low energy, high rate protons (400 MeV at ~2.2e15 protons/hr), with the beam available for about eight months a year, from November through June. MTest facilitates short-term projects with a proton beam of 120 GeV and secondary beams between 1-66 GeV, whereas MCenter caters to longer-term experiments with its secondary and tertiary beams down to 200 MeV.

Despite the extensive capabilities and heavy demand from significant projects like CMS, ATLAS upgrades, and various neutrino-related experiments, the FTBF faces challenges such as over-subscription by 10-20\%, aging infrastructure in the switchyard line, and substantial downtime that impacts experiment schedules. The increasing need for clean, low energy secondary beams and the potential discontinuation of the ITA in the absence of LINAC operations in the PIP-II era underscore the pressing requirements for facility updates and expansions to accommodate evolving research needs, including a high-intensity 800 MeV irradiation area for comprehensive particle exposure studies.

\paragraph{High Powered Targetry} This is discussed more as a part of Section~\ref{sec:accelsession}. 

\paragraph{Time Slicing of Neutrino Fluxes in Oscillation Experiments}

By utilizing a higher-frequency RF bunch structure for the primary proton beam on target and employing precision timing to select different energy and flavor spectra from a wide-band neutrino beam, based on the neutrinos' relative arrival times with respect to the RF bunch structure, we propose a 'stroboscopic' method. This approach is complementary to techniques that differentiate neutrino energy spectra based on their angle relative to the beam axis, allowing for the selection of varying energy spectra from the same on-axis detector and applying equally to both near and far detectors in an oscillation experiment. Discriminating energy and flavor of neutrinos produced by in-flight hadrons necessitates proton bunch lengths on the order of 100\,ps and comparable time resolution in the detector. The correlation of neutrino events with the parent proton interaction is currently hampered by the nanosecond-scale width of the proton bunches targeting the beam. By employing a superconducting RF cavity to rebunch the existing 53.1\,MHz RF bunch structure at a tenfold higher RF frequency, the requisite shorter bunch length can be achieved.

\paragraph{Fast tracking for triggering}

High-rate experiments at PIP-II / ACE could potentially enhance their physics capability significantly through the use of track triggers for signal identification, background rejection, and overcoming the challenges posed by sub-optimal beam timing structures. The success of this approach hinges on the deployment of sufficiently fast and granular tracking detectors. It's crucial to evaluate which experimental signatures, such as those from REDTOP and other experiments, would benefit the most from the incorporation of track triggers, thereby boosting the overall performance and outcome of these high-rate scientific endeavors.



\section{Physics, detector, and accelerator complementarity}
\label{sec:complementarity}
\subsection{CLFV - Dark Sectors}
\textit{Editors: Matt Solt, Jure Zupan}
\newline
The advantage of accelerator-based dark sector experiments over direct detection experiments is their relative insensitivity to specific dark matter models due to semi-relativistic versus non-relativistic scattering. Moreover, accelerator-based experiments can perform many type of searches and probe a variety of dark sector models at $pp$ collider, $e^+e^-$ colliders, proton and electron fixed target experiments, and beam dump experiments. One can search for visible final states, either by searching for promptly decaying resonances or displaced vertices, or for invisible final states through various kinematic handles: missing mass, missing momentum, and missing energy. The CLFV decays into light new physics states, should such states be found, would have a parametrically enhanced reach to high effective mass scales. For example, classic CLFV observables (such as $\mu \to e$, $\mu \to e \gamma$ and $\mu \to 3e$) probe scales at the level of $\sim 10^6$ GeV, the decay $\mu \to e a$, where $a$ is an axion, would probe a scale of $10^9$ GeV.

A clear outcome of the session was that the physics cases for performing CLFV measurements and dark sector searches is very strong, with much of the discussion devoted to outline R\&D direction to achieve the optimal physics program: 
\begin{itemize}
\item {\bf How do we maximize probing CLFV and dark sector searches with PIP-II and ACE capabilities?}
Through discussions with accelerator experts, a strong case was made to build a compressor ring in stages, build it as soon as possible, and make sure that both the AMF and the beam dump programs are possible. The preferred option would be to build a 2 GeV ring, but operate it first at 0.8 GeV. Accelerator physics considerations showed this should be possible, but the optimal scenario to operate the ring should be refined.
\item {\bf What type of beam?} Different experiments prefer different beam operation parameters, and the beam structure depends on how the dark sector is being sourced. Short bunches are probably preferred if it is produced in primary collisions, while the duration of the pulses are less important if it is from muon decays (e.g. from stopped muons).
\item {\bf Exotic signatures.}  The exotic signatures such as $\mu\to 5e$, $\mu \to e a$, ..., are interesting, but they are most motivated if they can be done already as part of an experiment aimed at some of the golden signatures (e.g., $\mu\to e \gamma$, $\mu\to e$ conversion, $\mu\to 3e$).
\item {\bf Muonium program.} There are other probes such as muonium-antimuonium oscillations, which are complementary to the rest of the program, since they probe different type of physics, though perhaps more exotic. 
\end{itemize}

\subsection{CLFV - Muon Collider}
\textit{Editors: Bob Bernstein, Sergo Jindariani, Diktys Stratakis}
\newline
Future Lepton Flavor Violation experiments (e.g. AMF and Mu2e-II) and the Muon Collider (MC) present a number of potential research and development (R\&D) opportunities, particularly in accelerator, target, and magnet technologies, that could complement each other and that was the main subject of the discussion. The following components were discussed: the Production Solenoid, the Target, the Compressor Ring, and the Fixed Field Alternating (FFA) Gradient synchrotron.
For each component, design parameters for Mu2e-II, AMF, and MC were laid out for commonalities and overlaps.  

After the discussion, it became evident that the Production Solenoid has some clear areas of overlap between Mu2e-II, AMF, and the MC. The study of megawatt-class targets is important for all experiments in the 800\,MeV to 8\,GeV range as AMF and the MC have similar requirements for target beam power, while Mu2e-II has lower requirements.  The protection of the superconducting solenoid will be important in all cases as well as the materials under consideration. As the designs of the production solenoids for each experiment evolve, it will be important to understand the synergies and potentially build a common team and test infrastructure for this R\&D.  

The proton Compressor Ring is another area of synergy although the requirements in energy and bunch length are about an order of magnitude different. Technologically, there are some synergies in the R\&D but significant differences in requirements make it unlikely that the same ring can be used for both facilities.  
For the muon storage/accelerator FFA, because of the large difference in the required energies, the designs do not have a lot of overlap and this is the least likely area of complementarity.

In general, the exploration between the MC and AMF/Mu2e-II offered some interesting directions for a collaborative R\&D program that could benefit both experimental thrusts. Shared tools and techniques such as simulation efforts can accelerate progress and cultivate joint expertise.

\subsection{CLFV - Neutrinos}
\textit{Editors: Innes Bigaran, Ryan Plestid, Anil Thapa}
\hspace{1cm}
\newline
Neutrinos are a unique probe of lepton flavor symmetries. Within the standard model, lepton flavor symmetries $L_e$, $L_\mu$ and $L_\tau$ are accidentally conserved: their conservation is not built into the theory but is emergent. There is no reason why some more complete theory should continue to preserve these symmetries. Indeed, the measurement of neutrino flavor oscillations (which provides evidence of nonzero neutrino masses) indicates that some physics beyond the renormalisable standard model must break these flavor symmetries. Neutrino masses in some extended theory could be either Dirac or Majorana in nature, where the latter could imply a violation of total lepton number $L=L_e+L_\mu+ L_\tau$. Moreover, charged leptons and neutrinos are linked by $SU(2)$ symmetry: as left-handed leptons, they are part of the same electroweak doublet, which leads to many neutrino mass models naturally giving rise to CLFV that can be searched for at experiments. 

Our discussion focused on highlighting how CLFV and neutrino experiments could mutually enrich the understanding of fundamental physics, particularly in light of the ACE upgrade. We separated the discussion into three stages: a discussion of CLFV experiments, of neutrino experiments, and complementarity between the two. What emerged as a key take-away message from our discussion was the need for closer interactions (a) between experimentalists in both CLFV and neutrinos, and (b) coordination between theory and experiment, on how best to move forward with united efforts to further explore this interplay. Some topics which were discussed in detail are listed below. 

\begin{itemize}
    \item \emph{Could we have neutrino-mass motivated targets for CLFV experiments?}\\
    CLFV processes involving the muon, i.e. $\mu\to e \gamma$, $\mu\to e$ conversion on nuclei, and muonium oscillation can be enhanced in various extensions to the SM that incorporate neutrino masses. Upgrades of experiments probing these processes are part of the ACE upgrade proposal, e.g. Mu2e-II, other CLFV muon decay probes. Models extended by right-handed neutrinos (e.g. type-I seesaw) can lead to large effects in CLFV if one explores the flavor structure of Dirac masses. Furthermore, extensions of the SM with exotic scalars (e.g. type-II seesaw) have a Yukawa coupling structure which correlates neutrino oscillation with CLFV decays. 
    \item \emph{Could the ACE upgrade make it viable for Fermilab to become a world-leader in the study of muon decays and muon conversion? }\\
    Developments of the advanced muon facility (AMF) mean that Fermilab will be a site where the sheer flux of muons produced will be beyond what has been seen elsewhere in the world. Given that this will be the case, then could we harness these muons to do all manners of study of their rare properties? There was broad consensus that provided the muon facility is designed to allow for efficient storage of both $\mu^+$ and $\mu^-$ then Fermilab is well positioned to offer the best experiments in all three golden channels of CLFV. 
    \item \emph{Is there a strong physics case for a tau optimized flux? }\\
    Neutrino oscillation experts broadly agreed that such a case is complementary rather than foundational. Statistics will be limited in tau samples, and atmospherics offer an existing probe of the $\nu_\tau$ sector with better statistical capabilities (and without the difficulties inherent to reconstructing $\tau^\pm$ leptons). 
    \item \emph{Could there be room for a neutrino factory-like concept at an AMF or a future muon collider? }\\
    An advantage of having neutrinos produced from muon decay rather than from meson decays is that they have a well-known energy and flavor spectrum.
    AMF with a muon storage ring could be extended to allow muons to decay and produce a focused beam of neutrinos, which would be essential for precision measurements of neutrino properties. Neutrino factories have historically been considered with a $\sim 1 ~{\rm GeV}$ muon storage ring; roughly commensurate with the needs for a muon collider. CLFV experiments prefer a much lower energy, the AMF plan is for a $\sim 20-30 ~{\rm MeV}$ storage ring with an alternate technology (a fixed field alternating gradient synchrotron). From the perspective of neutrino oscillation physics, the absolute energy scale is not crucial but rather sets the necessary baseline. The tight momentum resolution of the AMF storage ring would allow for a nearly mono-energetic neutrino source. Drawbacks include a smaller boost factor and therefore wider opening angle for the neutrino beam. The broad consensus was that this was worth further investigations. 
\end{itemize}

Although slightly beyond the scope of our discussion group, there was also interest expressed in a controlled neutrino source for hadronic physics studies, 
\begin{itemize}
    \item \emph{Could there be scope for neutrino and/or muon deep-inelastic scattering~(DIS) experiments in the ACE upgrade?}\\
    Muon DIS on polarized targets (an extension of Spin-Quest) could provide a fundamental understanding of the structure of nucleons. Polarized neutrino DIS experiments would open-up new opportunities to understand neutrino scattering cross sections, which are paramount for interpreting future results from experiments like DUNE. In general, these experiments can provide further information about the spin-structure of nucleons and allow for more precise measurements of neutrino properties. Moreover, deviations from SM predictions for neutrino-nucleon interactions could indicate the presence of new-physics effects. 
\end{itemize}

\subsection{Dark Sectors - Muon Colliders}
\textit{Editors: Cari Cesarotti, Yonatan Kahn}
\newline
The discussion on the complementarity between dark sector experiments and muon colliders proceeded in two stages:

\paragraph{How to leverage \emph{current} muon beams for dark sector searches.} We concluded that the requirements for muon beam cooling demonstrators are in some sense maximally orthogonal to the needs for dark sector experiments, which typically require high-intensity beams, a large integrated luminosity, and a regular bunch structure. However, we identified a promising possibility for a future experiment: using the discarded forward muons from Mu2e for a new beam dump experiment. The 8 GeV proton beam results in $\sim 10^{19}$ POT/year, integrating to about $10^{21}$ POT after PIP II in 2028--2029. Roughly 1\% of these protons will result in muons with energy of $\sim 3 \ {\rm GeV}$ or above, of which Mu2e only uses a negligible fraction. The time structure of the beam is not amenable to a missing momentum search, but a search for long-lived muonphilic particles with mass below $2m_\mu$ which decay to electrons or photons seems possible if a detector could be placed $\mathcal{O}(10 \ {\rm m})$ away. Investigating the feasibility of this setup would be an excellent synergistic use of the Fermilab accelerator facilities. While we are particularly interested in a muon beam for the ultimate muon collider, it is worth noting that the proton beam necessary for producing the muons could also be leveraged for auxiliary experiments as well.

\paragraph{How to leverage \emph{future} muon colliders for dark sector searches.} To date, muon collider physics studies have mostly focused on Higgs and electroweak precision phenomenology, as well as new particle with electroweak couplings. To our knowledge, a comprehensive study of weakly-coupled dark sectors (for example, the minimal dark photon, or a muon-specific dark force) at the collider itself has not yet been undertaken (though there have been studies of using the muon beam in a beam dump experiment). In addition, as was emphasized in Ian Low's talk, muon colliders may have access to yet another dark sector portal, the neutrino portal, due to the fact that neutrinos may be visible through their electroweak radiation in calorimeters. Investigating this unique feature of muon colliders may lead to new ideas for dark sector searches, for example those involving right-handed neutrinos, through a long-lived particle experiment analogous to FASER at the LHC. 

\subsection{Dark Sectors - Neutrinos}
\textit{Editors: Bhaskar Dutta, Alex Sousa, Jacob Zettlemoyer}
\newline
In the dark sector and neutrino discussion session, we considered a number of questions around the ACE plans and how the Fermilab neutrino and dark sector program may benefit from the ACE options. We note that complementarity exists within ACE to optimize the location and detection threshold of our detectors to probe a wide range of physics in the neutrino and dark sectors. There is obvious synergy between the detector technologies used for neutrino and dark sector physics. A constant theme of our discussion was the use of beam timing to increase the physics scope or neutrino and dark sector searches, either through coupling PIP-II to a proton accumulator ring, or using the nanosecond pulse structure of PIP-II operating in continuous wave (CW) mode. 

We also explored potential options to extend the reach of dark sector search experiments to unexplored regions of parameter space.  A PIP-II beam dump facility equipped with a high-Z target, and/or a higher beam energy, with judicious choices of detector location would offer excellent opportunities to probe new ranges of dark sector particle masses. Enhanced kaon production and detection would enable particularly sensitive searches. 

We discussed the broad ideas for producing and detecting dark sector particles and possible gaps in what we were considering. Some examples include light dark matter, axion-like particles, heavy neutral leptons, and millicharged particles. Gathering notable interest during the discussion, was the possibility of using resonant $\pi^0$ production at neutrino experiments for complementarity with the dark sector parameter search space using precision measurements of Dalitz decays or pion production. 

We discussed how the options under ACE would benefit the DUNE experiment, the current flagship neutrino experiment located in the US. The discussion around the physics scope expansion focused on the $\nu_{\tau}$ sector where an expansion of the beam power or energy beyond the baseline under ACE could enable significant inroads into improving our knowledge of this very difficult to measure sector. We noted that it may be possible under ACE to go beyond 2.4~MW, with the likely primary constraints being the target and absorber capabilities. Those constraints notwithstanding, reaching a beam power as high as 4~MW might be possible, with the limitation on the accelerator at that point becoming space charge effects. Additional discussion touched on the capability of the Main Injector to support such enhancements.

We discussed possible detector technologies that we could deploy at PIP-II under ACE for neutrino and dark sector searches. Some of these options are being explored in more detail and were discussed at a recent PIP-II Beam Dump Facility Workshop held at Fermilab in May 2023, with a white paper (Ref.~\cite{aguilararevalo2023physics}) summarizing the discussions in more detail. Interesting possibilities include water-based liquid scintillator (WbLS), granular detectors such as the LiquidO technology, fast tracking or fully pixelated detectors expanding upon the capabilities of liquid argon TPCs, optical liquid argon detectors with high photocathode coverage, enabling searches for keV-scale physics and expanding the range of dark sector models for which the detector is sensitive. Another possibility are CCD detectors performing powerful searches for millicharged particles. We considered possible synergies with the large tracking detectors used in collider experiments along with direct dark matter detection experiments in exploring how to improve detector development dedicated to searching for dark sector particles.

Further discussion centered on probes of new physics enabled by running the neutrino beam facilities in a beam dump mode, similar to how MiniBooNE operated previously, to pioneer proton beam dump based searches. The main advantage of running in this mode is the drastic reduction of the neutrino decay-in-flight and possible neutron backgrounds affecting new physics measurements by short-baseline detectors. We noted this is a possibility for the SBND detector once PIP-II comes online, making use of the BNB beamline in the PIP-II era, especially before the LBNF beam becomes operational. We note that a beam dump could be installed at the BNB to allow both on- and off-target running concurrently. This could provide a powerful tool for searching for light dark matter due to the detector distance from the target and the lower energy such that it covers complementary parameter space to other planned experiments. 

We also discussed other beam dump possibilities that are enabled by the ACE plan. 
Some experiments such as PIP2-BD can take advantage of the powerful timing capabilities of an $\mathcal{O}(1)$-GeV accumulator ring coupled to the PIP-II Linac. The Switchyard beamline could offer another potential location for a new beam dump facility, though significant investment would likely be required. 

There was an additional comment during the summary session about the increasing synergies of the neutrino and nuclear physics communities. Understanding neutrino-nucleus interactions is becoming essential to make precision measurements and predictions in the neutrino sector, while at the same time there is significant interest in studying nucleon and nuclear structure. One possible question in this direction is whether Fermilab under the ACE plan could provide something unique beyond the HEP community, with one example being nuclear physics measurements using neutrinos produced from a very intense GeV-scale muon beam. Such a beam, with high-precision knowledge of the muon flux, could enable detailed studies of neutrino-nucleus interactions.   

\subsection{Muon Collider - Neutrinos}
\textit{Editors: Christian Herwig, Zahra Tabrizi}
\newline
Projects such as the Deep Underground Neutrino Experiment (DUNE) and Hyper-Kamiokande (Hyper-K), now under construction, aim to achieve the statistical precision necessary to explore leptonic CP-invariance violation. These experiments, along with JUNO, are poised to conduct precision studies on various oscillation parameters, contributing to our understanding of neutrino physics and potentially shedding light on phenomena such as the MSW effect.

Neutrino factories, proposed around the time neutrino oscillations were discovered, offer a distinct approach to generating neutrino beams using muon decay and storage rings. Despite challenges in muon production and acceleration, these facilities present high luminosity and well-defined beam properties, making them ideal for studying neutrino oscillations. The synergy between neutrino factories and muon colliders is evident, as both face similar technological hurdles and stand to benefit from shared advancements.


Given the rich physics opportunities afforded by neutrino factories and the need to prepare for future neutrino physics programs beyond DUNE, detailed studies on their complex and detectors are timely. These efforts are crucial for advancing our understanding of fundamental physics and unlocking new discoveries in the realm of neutrino science.

The ACE upgrade presents various aspects potentially aligned with the needs of a muon collider, such as shared accumulator rings, though the extent of alignment depends on specific considerations. Discussions from other sessions, including S. Nagaitsev’s talk, provide further insights on these commonalities and may serve as a natural starting point to updated ACE-BR design proposals. However, any of the high-power proton-driven systems put forth as a part of the ACE-BR study can offer numerous opportunities for neutrino sources beyond decay-in-flight pions, including processes such as pion DAR, kaon decay, and muon decay.

Considering the post-DUNE/Hyper-K era, questions arise regarding the ideal experiment in light of anomalies persisting or dissipating. Should anomalies persist, a neutrino factory could offer valuable exploration avenues, while alternative scenarios prompt the consideration of other experimental avenues. 
In addition to this, a neutrino factory presents the compelling possibility to indirectly search for New Physics in a manner synergistic with high-energy probes through the effective field theory approach,
by matching WEFT and SMEFT parameters across energy scales.

In addition to the neutrino factory concept, high-energy neutrinos produced in MC collisions allow the possibility to conduct new measurements and searches through novel processes such as $\nu\bar\nu$ fusion processes.
Auxiliary MC detectors, akin to FASERNu, may facilitate investigations into additional new physics domains in addition to probing neutrino interactions at some of the highest energy scales. Furthermore, there is potential for novel measurements using high-energy neutrinos, with the physics case for such endeavors well-established since the inception of the neutrino factory concept. 
Exploration of topics such as Higgs-neutrino couplings and mapping the neutrino floor for dark matter detection are among the intriguing possibilities in this context.

\subsection{Accelerator technology and R\&D}
\textit{Editors: Jeff Eldred, Vladimir Shiltsev}
\label{sec:accelsession}

There were two talks at the Accelerator Parallel Session of the ACE Science Workshop providing technical overviews of the ACE Booster Replacement (ACE-BR) accelerator upgrade configurations. The first three ACE configurations involve a 2~GeV upgrade of the PIP-II Linac and a new 8~GeV RCS. The second three ACE configurations involve an 8-GeV upgrade of the PIP-II Linac and a new 8-GeV accumulator ring (AR). Overall the configurations concepts feature detailed technical design for many subsystems and have well-precedented projections of accelerator performance~\cite{nagaitsev2022rcs,belomestnykh2023linac}.

Since the time of the workshop, P5 has called for the development of a new strategic 20-year plan for Fermilab accelerator (Area Recommendation 12) oriented towards a future collider program (Recommendation 6). The ACE-BR scenarios were a detailed look at the proton accelerator complex upgrades with a core mission focus on the DUNE neutrino program. Consequently the exercise was able to identify R\&D items for accelerator development that are highly relevant to the P5 vision, but more work is needed to redevelop the specific upgrade scenarios towards incorporating future collider programs. More broadly, the critical need for advancements in accelerator R\&D is called out in P5 Recommendations 8-10.

The speakers were asked to identify technical risks and corresponding R\&D areas and raised a few topics. For both types of ACE-BR configurations, H$^{-}$ foil-stripping injection was identified the greatest technical challenge and most constraining performance limit. H$^{-}$ foil-injection requires managing particle loss scattering off the injection foil as well as mechanical degradation of the foil itself due to beam heating effects. Careful optimization of the H$^{-}$ foil injection section should proceed in parallel with development of H$^{-}$ laser-stripping technology (e.g. at Oak Ridge SNS~\cite{Cousineau2017laser}). Aside from injection, the ACE-BR configurations benefit from continuing innovations and improvements in SRF cavity and RF power technology, from Q-factors to accelerator gradients to achievable pulse length~\cite{belomestnykh2023linac,belomestnykh2022rf}. Lastly, two ``RCS'' ACE-BR configurations require use of metallized ceramic beampipes which would benefit from early prototyping (following J-PARC's precedent~\cite{Kinsho2005ceramic} but with new parameters and vendors).

One talk at the Accelerator Parallel Session addressed targetry development for the next-generation Intensity Frontier programs. The ACE Main Injector Ramp and Targetry (ACE-MIRT) plan has DUNE/LBNF beam powers increasing to multi-MW much faster than originally projected (ideally 2~MW around the 2033 possible end of Mu2e run), and in parallel envisions staging a series of neutrino (i.e. low-Z) targets rated for increasingly greater beam powers. Although the instantaneous thermal shock on the LBNF targets are not impacted by the increased repetition rate, the target lifetime will be impacted by the more rapidly accumulating radiation dose on the target materials. P5 called out the ACE-MIRT program in broad terms as part of Recommendation 2a.

In fact, high-power operations requires all targetry materials to perform at a higher accumulated dose; consequently the study of the performance of a variety of irradiated materials is a critical R\&D area. The Radiation Damage In Accelerator Target Environments (RaDIATE) collaboration has studied the performance of irradiated materials using the Brookhaven Linac Isotope Producer (BLIP) but there may be the opportunity to enhance Fermilab's proton irradiation capabilities using the PIP-II linac beam or the 8-GeV extracted beam \cite{Pellemoine2022rad,Ammigan2022target}. 

The ACE upgrade plan for the Fermilab proton complex envisions a diverse and growing science program at 2~GeV, 8~GeV, 120~GeV. Several of these programs may need targetry R\&D program programs of their own. In particular a significant synergy has been identified in the needs of the the Muon CLFV and Muon Collider programs (see Section 3.2 on the complementarity session). The programs require narrow high-Z targets for pion production in a large acceptance solenoid and it should be noted that Mu2e, Mu2e-II, AMF, and MuC proposals form a natural sequence of ever more ambitious targetry needs. To this end, the muon programs need not only the study of irradiated materials, but also the development and testing of target concepts.


One staging strategy for the AMF or MuC front-end could be to begin a target demonstrator facility at a lesser proton power as soon as possible, with a path to upgrade proton power, targetry, and physics capabilities as the facility matures. For AMF, this could involve the construction of a compact accumulator ring (also called proton compressor) to deliver 200-400~kW 20~ns proton pulses at 0.8 GeV. The program would serve a beam dump physics program~\cite{Toups:2022yxs} and a muon production target demonstrator initially, but compatible with 1+ MW beam power for AMF with a subsequent energy upgrade of the PIP-II linac. For the MuC front-end, this could involve a 8~GeV H$^{-}$ linac operating at 5mA for 2ms every 10~Hz into an 8~GeV AR (i.e. an ACE-BR configuration) for a 800~kW target demonstrator. Subsequently, the 8~GeV linac current would be upgraded to 10-20mA (1.6-3.2~MW), and a 8~GeV compressor ring and combiner beamline would be constructed to shorten the pulses from the 8-GeV AR for the MuC proton driver front-end.

One talk at the Accelerator Parallel Session specifically addressed immediate Muon Collider R\&D needs. P5 called for collider R\&D in Recommendation 4. In addition to the targetry development program, R\&D for high gradient RF in strong magnetic fields is needed for the cooling cell design. Within the next five years, the talk calls for a (1) conceptual design for a muon cooling demonstrator facility, (2) site identification and cost estimates for a muon cooling demonstrator facility, and (3) to begin fabrication of a prototype cooling cell for the facility. At Fermilab, the 8~GeV beamline and (ACE era) 2~GeV linac beamline are being investigated as locations to host the muon cooling demonstrator.

The final talk at the Accelerator Parallel Session discussed the PIP-II Accumulator Ring (PAR) proposal, a 0.8~GeV AR sited in between the PIP-II linac and the Fermilab Booster. The PAR ring would facilitate injection into the Fermilab Booster and provide a new PIP-II era experimental program. If the PAR ring could be constructed on a sufficiently aggressive timetable, it would provide excellent value in shortening the timetable for beam commissioning to LBNF. Regardless of timing, the PAR ring would concurrently be capable of a 100~kW low-duty factor proton program for beam dump physics~\cite{Toups:2022yxs}. However, the requirement of facilitating injection into the Booster (thereby benefiting the LBNF timeline) comes with tradeoffs for providing powerful beam to GeV-scale experiments. If the AR design had a more compact circumference (instead of matching Booster) it would deliver shorter pulses to experiments (improving the physics reach). Secondly, if the ring were not sited near the Booster, it could be compatible with a later energy upgrade when the PIP-II linac is extended (see staging discussion for AMF above). A conceptual design for an AR with this approach is under development.


Not directly discussed in the Accelerator Parallel Session were the requirements of optimizing the reliability of the Fermilab Booster. Later, P5 highlighted the need for the assessment of Booster reliability and the identification of pro-active measures to guarantee the longevity of Booster operations. On March 5th, 2024 a mini-workshop ``AD Prep for DUNE PIP-II Era Workshop'' will be held as a preliminary measure towards addressing this important topic. Whether any new upgrades for Booster reliability should be incorporated into ACE-MIRT or considered as standalone items is also a matter to be determined on a case-by-case basis.

\section*{Acknowledgments}

We would like to thank all the speakers of the workshop for their time and expertise in preparing and giving their presentations.  In particular, we thank Sergey Belomestnykh, David Sperka, Ishara Fernando, Evan Niner, Jaehoon Yu for their feedback on this report. We would also like to acknowledge all of the attendees of the workshop for their comments and contributions which provided valuable input and lively discussion.

\bibliography{references}
\end{document}